\documentclass[12pt]{article}
\usepackage{amsfonts}
\usepackage{amssymb}
\usepackage{color}
\usepackage{graphicx}

\newcommand{\be}{\begin{equation}}
\newcommand{\en}{\end{equation}}
\newcommand{\bea}{\begin{eqnarray}}
\newcommand{\ena}{\end{eqnarray}}
\newcommand{\beano}{\begin{eqnarray*}}
\newcommand{\enano}{\end{eqnarray*}}

\catcode `\@=11 \@addtoreset{equation}{section}

\catcode `\@=12
\begin{document}

\begin{center}
{\Large \textbf{The role of information in a two-traders market}} \vspace{2cm%
}\\[0pt]

{\large F. Bagarello}
%\footnote[1]{ Dipartimento di Matematica ed Applicazioni,
%Fac.Ingegneria, Universit\`a di Palermo, I - 90128  Palermo, Italy}
\vspace{3mm}\\[0pt]
DEIM, Facolt\`{a} di Ingegneria,\\[0pt]
Universit\`{a} di Palermo, I - 90128 Palermo, Italy\\[0pt]
E-mail: fabio.bagarello@unipa.it\\[0pt]
home page: www.unipa.it/fabio.bagarello \vspace{8mm}\\[0pt]
{\large E. Haven}
%\footnote[1]{ Dipartimento di Matematica ed Applicazioni,
%Fac.Ingegneria, Universit\`a di Palermo, I - 90128  Palermo, Italy}
\vspace{3mm}\\[0pt]
School of Management {and Institute of Finance},\\[0pt]
University of Leicester,\\[0pt]
Leicester, United Kingdom\\[0pt]
E-mail: e.haven@le.ac.uk \vspace{2mm}\\[0pt]
\end{center}

\vspace*{2cm}

\begin{abstract}
\noindent In a very simple stock market, made by only two \emph{initially
equivalent} traders, we discuss how the information can affect the
performance of the traders. More in detail, we first consider how the
portfolios of the traders evolve in time when the market is \emph{closed}.
After that, we discuss two models in which an interaction with the outer
world is allowed. We show that, in this case, the two traders behave
differently, depending on \textbf{i)} the amount of information which they
receive from outside; and \textbf{ii) }the quality of this information.
\end{abstract}

\vfill

\newpage

% Section 1

\section{Introduction and motivations}

In a series of papers, \cite{bag1}-\cite{bag4}, one of us (FB) has shown how
the Heisenberg time evolution used for quantum mechanical systems can be
adopted in the analysis of some simplified stock markets. After these
original applications, the same tools were also used for rather different
macroscopic systems. A recent monograph on these topics is \cite{bagbook}.
In the cited papers and in \cite{bagbook} the role of information was, in a
certain sense, only incorporated by properly choosing some of the constants
defining the Hamiltonian of the system we were considering.

On the other hand, the other author (EH), {following the original idea of
\cite{khren1}, }considered the role of information for stock markets, \cite%
{hav1}-\cite{hav2}, mainly adopting the Bohm view to quantum mechanics,
where the information is carried by a pilot wave function $\Psi (x,t)$,
satisfying a Schr\"{o}dinger equation of motion, and which, with simple
computations, produces what is called \emph{a mental force} which has to be
added to the other \emph{hard} forces acting on the system, producing a full
Newton-like classical differential equation.

In this paper we try first to incorporate the effect of this mental force at
a purely quantum mechanical level. After that, we consider a simplified
stock market, which, to simplify the notation, we consider with just two
traders $\tau _{1}$ and $\tau _{2}$, describing what happens \textbf{before}
the trading begins, i.e. in the phase in which the information begins to
circulate in the market, and is used by the traders to decide their next
moves. {The rationale for focusing on the way information can influence
valuation of portfolios is a very important topic in finance and economics.
We stress that it is the modeling of the information which is at the heart
of the problem in such valuation exercises. We believe this paper shows that
tools from quantum mechanics can aid in a very valuable way to this modeling
challenge. }

It may be worth stressing that our analysis continues a nowadays rather rich
literature on the role of quantum mechanics in economics, see \cite{accbouk}-%
\cite{piotr} for instance, which shows that an increasing number of
researchers believe that some of the aspects of a \emph{real} stock market
could be described by adopting tools and ideas coming from quantum
mechanics. {We should stress that, in our knowledge, the first paper
where such a connection between quantum mechanics and finance appeared is
\cite{seg}, where the authors suggested that non commuting operators are
really needed in the description of a realistic market to prevent exact
knowledge of the price of a share and of its forward time derivative. See
also \cite{baa}. These two quantities, in \cite{baa} and \cite{seg}, were
associated to operators having the same commutation rule as the position and
the momentum operators in ordinary quantum mechanics, and therefore obey an
uncertainty principle. Furthermore, there is scope to argue that for
instance the central concept of non-arbitrage in finance has connections
with hermiticity in quantum mechanics. Baaquie \cite{baa} has shown that the
hamiltonian of the Black-Scholes equation is not hermitian. This
non-existence of hermiticity is narrowly related to the absence of arbitrage
(the existence of a martingale). Clearly, hermiticity on itself is not
making anything quantum mechanical as such, but it is still an important
argument. There are other interesting arguments, such as the way hidden
variable theory can connect with the (non-observable) state prices, in
again, the non-arbitrage theorem. See \cite{manu} Finally, we also want to
mention that in the context of decision theory, notably in the resolving of
some expected utility paradoxes, the use of quantum probability is very
promising. Those paradoxes lie at the base of many economics/finance models.
We document those achievements in \cite{manu}. In essence, the use of
quantum mechanical techniques into social science revolve really around
formalizing information. See \cite{bagbook}}.

The paper is organized as follows: in the next section we briefly discuss
how the pilot wave function can be incorporated in our Heisenberg-like
dynamics. Then, in Section III we introduce a first model of a closed
market, where the information (or, in our setting, the \emph{lack of
information}, LoI in the following) will behave as the other operators,
i.e., it will be described by ordinary two-modes bosonic operators. In
Section IV we replace these operators with two families of bosonic
operators, describing sources and sinks of information which modify, in the
way described below, directly the portfolios of the traders. In Section V,
finally, we consider a more complete model where the outer world contributes
in the definition of the strategies of the traders in a more realistic way,
i.e. by contributing to the information of the traders, rather than {being
the information} by itself. Section VI contains our conclusions.

\section{Some preliminaries}

In FB's approach to stock markets the maybe crucial ingredient of the model
is the Hamiltonian operator $H$ which is taken to describe the system. In
\cite{bagbook} several useful rules have been proposed to fix the expression
of $H$. We need now to incorporate in $H$ the effect described by the pilot
wave function, extending, for instance, what is discussed in \cite{chou}. {%
See also \cite{khren1}}. Let us recall here the essential steps: the main
ingredient is the (two-dimensional, in our case) pilot wave function, $\Psi
(q_{1},q_{2})$, which evolves in time according to the Schr\"{o}dinger
equation of motion
\[
i\frac{\partial \Psi (q_{1},q_{2};t)}{\partial t}=\left[ -\frac{\hbar ^{2}}{%
2m}\sum_{j=1}^{2}\frac{\partial ^{2}}{\partial q_{j}^{2}}+V(q_{1},q_{2})%
\right] \Psi (q_{1},q_{2};t),
\]%
where $\hbar $ and $m$ have a suitable economics based meaning\footnote{{It
is to be noted that to give an economics based interpretation of }$\hbar ${\
is still a very difficult challenge.}}, \cite{khren1} and \cite{chou}, and $%
V(q_{1},q_{2})$ is the potential due to the hard economics based effects.
Then, calling $R(q_{1},q_{2})=|\Psi (q_{1},q_{2})|$, a new potential is
constructed by defining $U(q_{1},q_{2})=-\frac{1}{R(q_{1},q_{2})}%
\sum_{j=1}^{2}\frac{\partial ^{2}R(q_{1},q_{2})}{\partial q_{j}^{2}}$, and $%
U(q_{1},q_{2})$ produces the mental forces affecting the traders: $%
g_{j}(q_{1},q_{2})=-\frac{\partial U(q_{1},q_{2})}{\partial q_{j}}$, $j=1,2$%
. {Please note the definition of this new potential is not foreign to
physics but is squarely steeped into Bohmian mechanics (which is a
particular interpretation of quantum mechanics). The key references are \cite%
{bohm1} and \cite{bohm2}. }Finally, if we call $\pi _{j}(t)$ the value of
the portfolio\footnote{%
This approach is slightly different from \cite{khren1,chou}, but it is more
natural in the present context.} of $\tau _{j}$, its time evolution is
driven by the following classical ({Newtonian-like) }differential equation:
\[
\dot{\pi}_{j}(t)=-\frac{\partial V(q_{1},q_{2})}{\partial q_{j}}-\frac{%
\partial U(q_{1},q_{2})}{\partial q_{j}}%
=:f_{j}(q_{1},q_{2})+g_{j}(q_{1},q_{2}),
\]%
$j=1,2$, with obvious notation. Hence, the time evolution of $\pi _{j}(t)$
is governed by hard factors ($f_{j}$) as well as by the \emph{financial
mental force} $g_{j}$, \cite{khren1,chou}.

\vspace{2mm}

What is important for us is the potential $U(q_{1},q_{2})$, which
represents, in some sense, the fact that $\tau _{1}$ and $\tau _{2}$ are
reached by two, in general different, amounts of information. {%
Therefore it is natural} to assume that $%
U(q_{1},q_{2})=U_{1}(q_{1})+U_{2}(q_{2})$, with $U_{1}$ and $U_{2}$, in
general, different functions of their arguments. In this way we can model,
quite simply, the fact that $g_{1}=-\frac{\partial U}{\partial q_{1}}$ can
be different from $g_{2}-\frac{\partial U}{\partial q_{2}}$ and, quite
importantly from a technical point of view, the quantum-like Hamiltonian
constructed out of this potential can be viewed as the sum of two one-body
Hamiltonians, \cite{rom}. Just to fix the ideas, the Hamiltonian for the
market will contain a contribution like this:
\[
\Omega _{1}\,i_{1}^{\dagger }\,i_{1}+\Omega _{2}\,i_{2}^{\dagger }\,i_{2},
\]%
for closed systems, or having a slightly more general expression for open
systems. Here $\Omega _{1}$ and $\Omega _{2}$ are positive numbers, while $%
i_{j}$'s are bosonic operators (i.e. $[i_{j},i_{k}^{\dagger }]=1\!\!1\delta
_{j,k}$). This is exactly the kind of contribution one has for a two
particle systems in ordinary quantum theory, when the free energies of the
particles are expected to be different.

\vspace{2mm}

\textbf{Remark:--} It should be stressed that when we {use above and in the
sequel of this paper the terms }\emph{closed} or \emph{open} systems, this
terminology should be taken with a certain care. In fact, we call a system
\textbf{closed } when the information is described by a two-modes bosonic
operator, obeying the commutation rules above. In other words, information,
cash and shares are operators exactly \emph{of the same kind}. This will be
made more explicit in the next sections. However, since we expect the
information comes from outside the market, it would probably be more
appropriate to speak of \emph{absence of reservoir}.

\vspace{2mm}

The above remark is related to another interesting aspect of the models
proposed here, which somehow look different from those considered in \cite%
{bag1}-\cite{bag4}. In these former papers, the cash and the number of
shares of the traders were assumed to be constant in time: the shares are
not created or destroyed, for instance. Here, on the contrary, we allow for
such a possibility, so that bankruptcy can be discussed within our present
scheme. Moreover, we are not even assuming that the cash is only used to buy
shares, so that it needs not to be preserved in time, as well. However, we
will see in the next sections that other observables will be constant, and
we will see that these observables do have a clear economical meaning,
indeed.

As already anticipated, in this paper we will be essentially interested {not
in the interaction between the traders, but rather in }the effect of the
outer world in \emph{preparing the system}, {i.e. in fixing the
initial status of the various traders after they have been reached by the
information but before they start to trade.} For this reason, if we call $%
H_{full}=H+H_{ex}$ the Hamiltonian describing the traders and the
information, and if with $H_{ex}$ we {mean} that part of $H_{full}$
describing the exchanges between $\tau _{1}$ and $\tau _{2}$, see \cite%
{bagbook}, we will only be interested here in $H$. It is like if we are
considering two different time intervals: in the first one, $[0,t_{1}]$, the
two traders, which are indistinguishable at $t=0$, receive a different
amount of information. This allows them to react in different ways, so that,
at time $t_{1}$, they are expected to become different. In this interval, $%
H_{full}$ coincides with $H$. For $t>t_{1}$, the two traders have been \emph{%
prepared} in different ways, and the Hamiltonian is now $H_{ex}$ (plus, in
general, a free contribution). In other words, we could think {of }writing $%
H_{full}=H+\Theta (t-t_{1})\,H_{ex}$, where $\Theta (t)=1$ if $t>0$, while $%
\Theta (t)=0$ otherwise. Since, {in this paper, }we will only {be }%
interested in the first time interval, $[0,t_{1}]$, the role of $H_{ex}$
will not be very relevant here. We will say more on $H_{ex}$ in our
conclusions. This approach has also a quite useful technical consequence:
there is no real need, at this stage, to introduce the price of the share
and to consider its dynamical behavior. This becomes really important, of
course, when transactions are considered, not before. For this reason, in
this paper, the price of the shares (just a single kind of shares!) will be
fixed to be one. Again, we will say more on this in Section VI.

\section{A first model with no reservoir}

The first model we want to consider is described by the following
Hamiltonian:
\begin{equation}
\left\{
\begin{array}{ll}
H=H_{0}+H_{inf}, &  \\
H_{0}=\sum_{j=1}^{2}(\omega _{j}^{s}\hat{S}_{j}+\omega _{j}^{c}\hat{K}%
_{j}+\Omega _{j}\hat{I}_{j}), &  \\
H_{inf}=\lambda _{inf}\sum_{j=1}^{2}\left( i_{j}(s_{j}^{\dagger
}+c_{j}^{\dagger })+i_{j}^{\dagger }(s_{j}+c_{j})\right) .\label{31} &
\end{array}%
\right.
\end{equation}

Here $\hat S_j:=s_j^\dagger s_j$, $\hat K_j:=c_j^\dagger c_j$, and $\hat
I_j:=i_j^\dagger i_j$, $j=1,2$. The following canonical commutation
relations (CCRs) are assumed:
\begin{equation}
[s_j,s_k^\dagger]=[c_j,c_k^\dagger]=[i_j,i_k^\dagger]=\delta_{j,k}1 \!\! 1,
\label{32}
\end{equation}
where $1 \!\! 1$ is the identity operator. All the other commutators are
zero.

The meaning of these operators is widely discussed in \cite{bagbook}: $s_{j}$
destroys a share in the portfolio of $\tau _{j}$, see below, while $%
s_{j}^{\dagger }$ creates a share. The operators $c_{j}$ and $c_{j}^{\dagger
}$ respectively lower and rise the amount of cash of $\tau _{j}$. Finally, $%
i_{j}^{\dagger }$ increases the LoI of $\tau _{j}$, while $i_{j}$ decreases
it\footnote{%
Although it is important to stress the relation between LoI and entropy, we
do not take it up in this paper. We thank one of the referees for pointing
this out. There exists an interesting relationship between the average
quantum potential and Fisher information. This was proposed in \cite{reg}.
See also \cite{hawkins}.}. Therefore, the meaning of $H_{inf}$ is the
following: whenever the LoI decreases (because of $i_{j}$), the value of the
portfolio operator of $\tau _{j}$, $\hat{\pi}_{j}:=\hat{S}_{j}+\hat{K}_{j}$
\footnote{%
Observe that, since the price of the share is one, this is the sum of the
cash of $\tau _{j}$ and the value of his shares.}, increases (because of $%
s_{j}^{\dagger }+c_{j}^{\dagger }$). Of course, since $H_{inf}$ also
contains the adjoint contribution $i_{j}^{\dagger }(s_{j}+c_{j})$, if the
LoI increases, then $\hat{\pi}_{j}$ decreases.

It is not hard to check that, calling $\hat{M}_{j}:=\hat{S}_{j}+\hat{K}_{j}+%
\hat{I}_{j}=\hat{\pi}_{j}+\hat{I}_{j}$, the following is true: $[H,\hat{M}%
_{j}]=0$, $j=,1,2$. Consequently, even {if} the cash and the shares are not
separately preserved, the sum of the portfolio and the LoI of each trader
(and therefore of the whole market) stays constant. This has an economical
meaning: whenever the LoI increases, it is natural to imagine that the value
of the portfolio of the related trader should decrease, while having more
information means having more chances to increase {one's wealth}. And this
is exactly what the commutativity between $H$ and $\hat{M}_{j}$ implies.

The equation of motion for $\tau_j$ can be easily deduced using the
Heisenberg equation of motion $\dot X(t)=i[H,X(t)]$. We find that
\begin{equation}
\dot X_j(t)=-iT_jX_j(t),  \label{33}
\end{equation}
where
\begin{equation}
X_j(t)=\left(
\begin{array}{c}
s_j(t) \\
c_j(t) \\
i_j(t) \\
\end{array}
\right),\qquad T_j=\left(
\begin{array}{ccc}
\omega_j^s & 0 & \lambda_{inf} \\
0 & \omega_j^c & \lambda_{inf} \\
\lambda_{inf} & \lambda_{inf} & \Omega_j \\
&  &
\end{array}
\right).  \label{33bis}
\end{equation}
The solution can be written as $X_j(t)=V_j(t)X_j(0)$, where $%
V_j(t)=U_j\Sigma_j(t)U_j^{-1}$, $U_j$ being the matrix which diagonalizes $%
T_j$, $U_j^{-1}T_jU_j=diag\{\sigma_1^{(j)},\sigma_2^{(j)},\sigma_3^{(j)}\}=:%
\sigma_j$, and $\Sigma_j(t)=\exp\{-i\,\sigma_j\,t\}=\left(
\begin{array}{ccc}
e^{-i\sigma_1^{(j)}t} & 0 & 0 \\
0 & e^{-i\sigma_2^{(j)}t} & 0 \\
0 & 0 & e^{-i\sigma_3^{(j)}t}%
\end{array}
\right)$.

The state of the system, at $t=0$, is assumed to be $\varphi_\mathcal{G}%
:=\varphi_{S_1,K_1,I_1,S_2,K_2,I_2}$, which, see \cite{bagbook}, can be
constructed by a vacuum $\varphi_{\mathbf{0}}$, $c_j\varphi_{\mathbf{0}%
}=s_j\varphi_{\mathbf{0}}=i_j\varphi_{\mathbf{0}}=0$, $j=1,2$, acting with
powers of the raising operators $c_j^\dagger$, $s_j^\dagger$ and $%
i_j^\dagger $, and normalizing the result. The vector $\varphi_\mathcal{G}$
describes a market in which, at $t=0$, $\tau_1$ possesses $S_1$ shares, $K_1$
units of cash, and is affected by a LoI equal to $I_1$. Similarly for $%
\tau_2 $. The time evolution of the cash of $\tau_j$ is deduced by computing
$N_{K_j}(t):=\left<\varphi_\mathcal{G}, c_j^\dagger(t)c_j(t)\varphi_\mathcal{%
G}\right>$. Analogously, the number of shares are given by $%
N_{S_j}(t):=\left<\varphi_\mathcal{G}, s_j^\dagger(t)s_j(t)\varphi_\mathcal{G%
}\right>$. The value of the portfolio of $\tau_j$ is just the sum of $%
N_{K_j}(t)$ and $N_{S_j}(t)$:
\begin{equation}
\pi_j(t)=\left<\varphi_\mathcal{G}, \hat\pi_j(t)\varphi_\mathcal{G}%
\right>=N_{S_j}(t)+N_{K_j}(t).  \label{34}
\end{equation}

\begin{figure}[h]
\begin{center}
\includegraphics[width=0.47\textwidth]{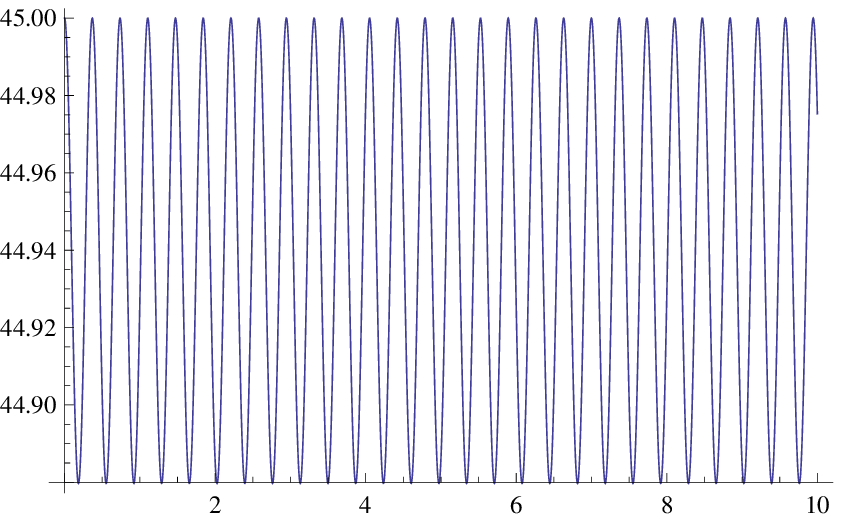}\hspace{8mm} %
\includegraphics[width=0.47\textwidth] {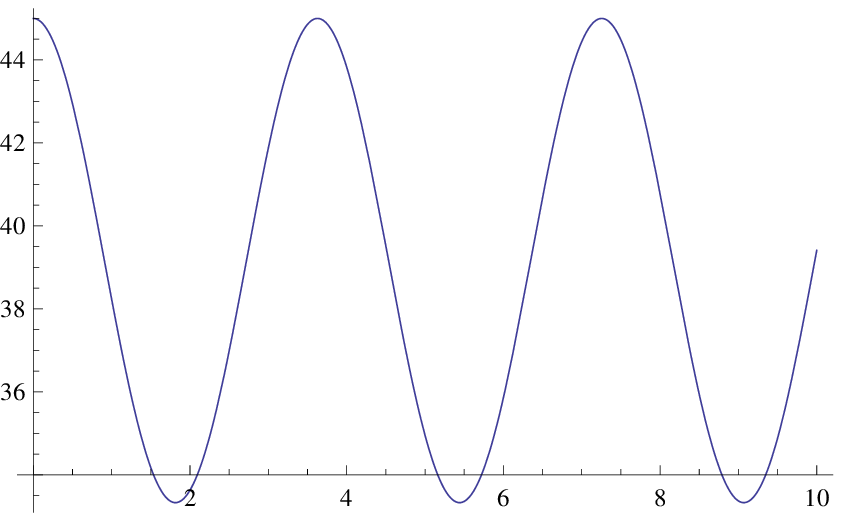}\hfill\\[0pt]
\end{center}
\caption{{\protect\footnotesize $\protect\pi(t)$ for $\protect\omega^s=%
\protect\omega^c=20$ (left) and $\protect\omega^s=\protect\omega^c=2$
(right) }}
\label{fig1}
\end{figure}

In the previous analysis carried out by one of us (FB), it was suggested
that the parameters of the free Hamiltonians influence significantly the
interacting system, while they play no role if no interaction occurs. The
same conclusion also follows from the analysis carried {out} here. To put in
evidence this aspect, it is better to choose $\tau _{1}$ completely \emph{%
equivalent} to $\tau _{2}$: hence we fix, first of all, $S_{1}=S_{2}$, $%
K_{1}=K_{2}$, $I_{1}=I_{2}$. This means that the initial conditions of the
two traders are identical. Moreover, we also ask that $\lambda _{inf}=0.5$
(just to fix the ideas) and that $\omega _{1}^{s}=\omega _{2}^{s}=:\omega
^{s}$, $\omega _{1}^{c}=\omega _{2}^{c}=:\omega ^{c}$ and $\Omega
_{1}=\Omega _{2}$. Therefore, the Hamiltonian for $\tau _{1}$ is identical
to the one for $\tau _{2}$. This implies that the matrix $T_j$ in (\ref%
{33bis}) for the two traders are equal: $T_{1}=T_{2}$ and, clearly, the
portfolios of the two traders coincide during their time evolution: $\pi
_{1}(t)=\pi _{2}(t)=:\pi (t)$. What appears interesting to us is that the
higher the values of $\omega ^{s}$ and $\omega ^{c}$, the smaller the
amplitude of the oscillations of the portfolios: as in very different
systems, \cite{bagbook}, in this simple situation, the parameters of the
free Hamiltonian behave as a sort of inertia for the traders, restricting
more and more the widths of the oscillations. In Figure \ref{fig1} we plot $%
\pi (t)$ for $S_{1}=30$, $K_{1}=15$ and $I_{1}=5$, and for $\omega
^{s}=\omega ^{c}=20$, $\Omega _{1}=\Omega _{2}=3$ (left), and for $\omega
^{s}=\omega ^{c}=2$, $\Omega _{1}=\Omega _{2}=3$ (right). We see in both
cases oscillation of $\pi (t)$, but the range (and the frequencies) of the
oscillations are quite different in the two cases.

\begin{figure}[h]
\begin{center}
\includegraphics[width=0.47\textwidth]{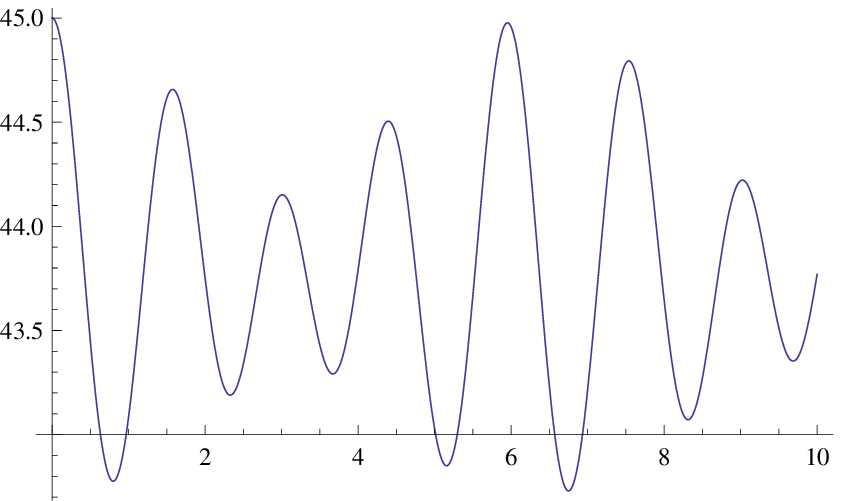}\hspace{8mm%
} \includegraphics[width=0.47\textwidth] {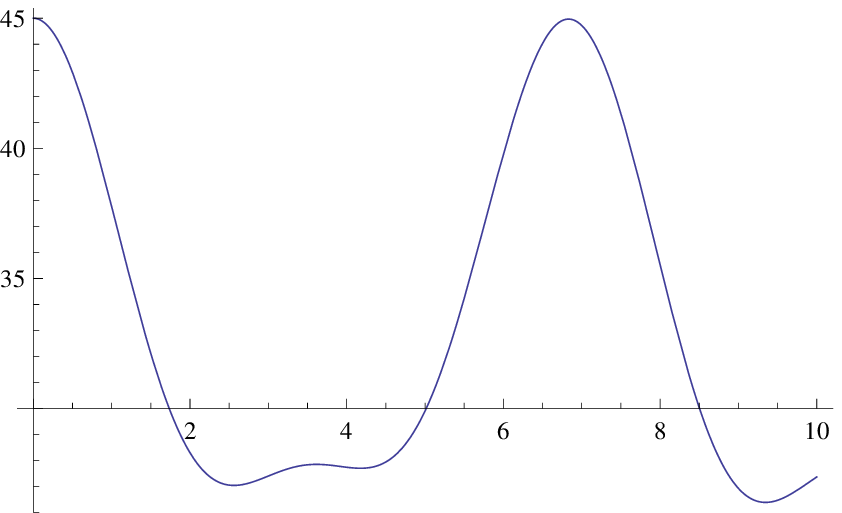}\hfill\\%
[0pt]
\end{center}
\caption{{\protect\footnotesize $\protect\pi_1(t)$ (left) and $\protect\pi%
_2(t)$ (right) for $\protect\omega^s=1$, $\protect\omega^c=2$, $\Omega_1=5$
and $\Omega_2=1$ }}
\label{fig2}
\end{figure}

\begin{figure}[h]
\begin{center}
\includegraphics[width=0.47\textwidth]{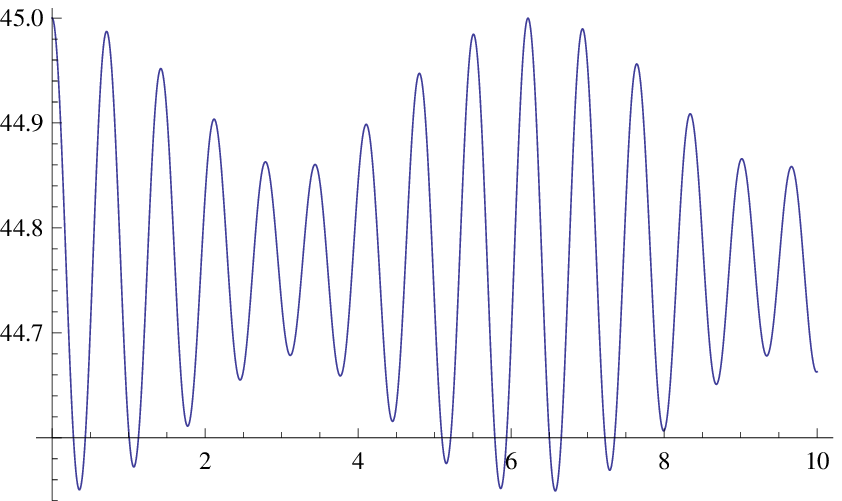}\hspace{8mm%
} \includegraphics[width=0.47\textwidth] {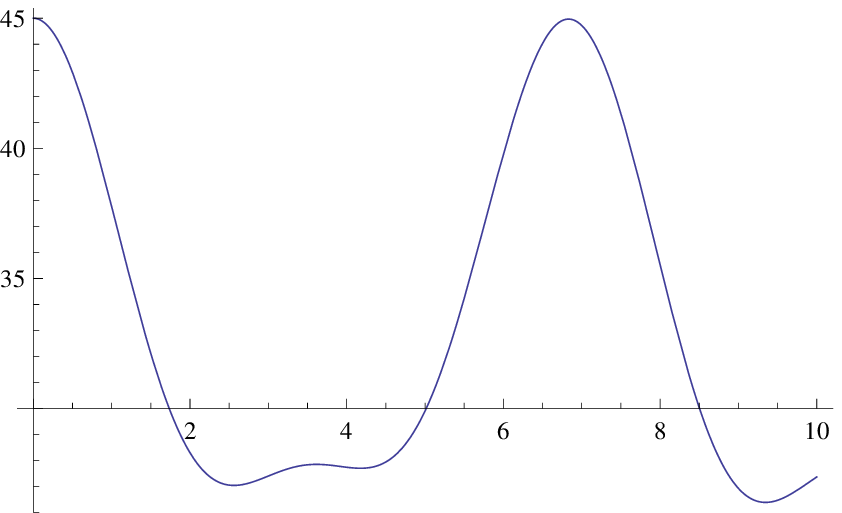}\hfill\\%
[0pt]
\end{center}
\caption{{\protect\footnotesize $\protect\pi_1(t)$ (left) and $\protect\pi%
_2(t)$ (right) for $\protect\omega^s=1$, $\protect\omega^c=2$, $\Omega_1=10$
and $\Omega_2=1$ }}
\label{fig3}
\end{figure}

Let us now consider the case in which the two traders, originally (i.e. at $%
t=0$) \emph{prepared} in the same way ($S_{1}=S_{2}=30$, $K_{1}=K_{2}=15$, $%
I_{1}=I_{2}=5$), are no longer completely equivalent: again we put $\omega
_{1}^{s}=\omega _{2}^{s}=:\omega ^{s}$ and $\omega _{1}^{c}=\omega
_{2}^{c}=:\omega ^{c}$, but we now assume that $\Omega _{1}>\Omega _{2}$. In
particular, in Figure \ref{fig2} we plot $\pi _{1}(t)$ (left) and $\pi
_{2}(t)$ (right) for the choices $\omega ^{s}=1$, $\omega ^{c}=2$, $\Omega
_{1}=5$ and $\Omega _{2}=1$. In Figure \ref{fig3} the parameters are $\omega
^{s}=1$, $\omega ^{c}=2$, $\Omega _{1}=10$ and $\Omega _{2}=1$.

We see again that, increasing $\Omega _{1}$ produces a smaller amplitude of
oscillation (bigger inertia) and a larger frequency. In fact, it is also
evident from Figures \ref{fig1}-\ref{fig3} that the omega's affect the
(pseudo-)frequencies of the functions $\pi_j(t)$: it seems that a larger
loss of information induces more frequent changes in portfolio values.

From both figures it is evident how the values of the free Hamiltonian do in
fact play a relevant role in the time evolution of the interacting system.
This is interesting since, if we take $\lambda _{inf}=0$, then both $\pi
_{1}(t)$ and $\pi _{2}(t)$ turn out to stay constant in time: no information
$\Rightarrow $ no action!

Rather than considering other aspects of this model, we consider now a
different, and more interesting Hamiltonian, based on the idea that the LoI
is related to the outer world surrounding the traders (the rumors, the news,
facts, etc.).

\section{The reservoir \protect\underline{is} the information}

The Hamiltonian we are interested in here is simply a generalized version of
that introduced in the previous section. The main difference is that the two
pairs of \emph{lack-of-information operators}, $(i_{1},i_{1}^{\dagger })$
and $(i_{2},i_{2}^{\dagger })$, are replaced by two families of similar
operators, labeled by the real numbers: $(i_{1}(k),i_{1}^{\dagger }(k))$ and
$(i_{2}(k),i_{2}^{\dagger }(k))$, where $k\in \mathbb{R}$ could be viewed as
a wave number, satisfying the commutation rules
\[
\lbrack i_{n}(k),i_{m}^{\dagger }(q)]=\delta _{n,m}\delta (k-q)1\!\!1,
\]%
all the other commutators being zero. The Hamiltonian is now

\begin{equation}
\left\{
\begin{array}{ll}
H=H_{0}+H_{inf}, &  \\
H_{0}=\sum_{j=1}^{2}(\omega _{j}^{s}\hat{S}_{j}+\omega _{j}^{c}\hat{K}%
_{j}+\int_{\mathbb{R}}\Omega _{j}(k)\hat{I}_{j}(k)\,dk), &  \\
H_{inf}=\lambda _{inf}\sum_{j=1}^{2}\int_{\mathbb{R}}\left(
i_{j}(k)(s_{j}^{\dagger }+c_{j}^{\dagger })+i_{j}^{\dagger
}(k)(s_{j}+c_{j})\right) \,dk.\label{41} &
\end{array}%
\right.
\end{equation}%
Once again, the model admits some integrals of motion: $\hat{M}_{j}:=\hat{S}%
_{j}+\hat{K}_{j}+\hat{I}_{j}=\hat{\pi}_{j}+\hat{I}_{j}$, where $\hat{\pi}%
_{j}=\hat{S}_{j}+\hat{K}_{j}$ is, as before, the portfolio operator for $%
\tau _{j}$ and $\hat{I}_{j}=\int_{\mathbb{R}}\hat{I}_{j}(k)\,dk$ is its
\emph{full LoI}. The existence of these integrals of motion have the same
economical meaning we have already discussed in the previous section, and {%
this }will not be repeated here.

The Heisenberg equations of motion for the operators of $\tau_j$ can be
easily found:

\begin{equation}
\left\{
\begin{array}{ll}
\frac{d}{dt}s_{j}(t)=-i\omega _{j}^{s}s_{j}(t)-i\lambda _{inf}\int_{\mathbb{R%
}}i_{j}(k,t)\,dk, &  \\
\frac{d}{dt}c_{j}(t)=-i\omega _{j}^{c}c_{j}(t)-i\lambda _{inf}\int_{\mathbb{R%
}}i_{j}(k,t)\,dk, &  \\
\frac{d}{dt}i_{j}(k,t)=-i\Omega _{j}(k)i_{j}(k,t)-i\lambda _{inf}(s(t)+c(t)).%
\label{42} &
\end{array}%
\right.
\end{equation}%
We will solve this system under the simplifying assumption that $\omega
_{j}^{s}=\omega _{j}^{c}=:\omega _{j}$. This is technically convenient,
since in this case the system above can be replaced by the simpler set of
equations
\begin{equation}
\left\{
\begin{array}{ll}
\frac{d}{dt}c_{j}(t)=-i\omega _{j}c_{j}(t)-i\lambda _{inf}\int_{\mathbb{R}%
}i_{j}(k,t)\,dk, &  \\
\frac{d}{dt}i_{j}(k,t)=-i\Omega _{j}(k)i_{j}(k,t)-i\lambda
_{inf}(2c_{j}(t)+e^{-i\omega _{j}t}(s_{j}(o)-c_{j}(0))).\label{43} &
\end{array}%
\right.
\end{equation}%
It is well known how to proceed in this case, \cite{bagbook}: we first
rewrite the second equation in integral form, and then we replace this
formula in the first equation above. Now, assuming that $\Omega
_{j}(k)=\Omega _{j}\,k$ for some positive $\Omega _{j}$, and recalling that $%
\int_{\mathbb{R}}e^{-i\Omega _{j}\,k(t-t_{1})}\,dk=\frac{2\pi }{\Omega _{j}}%
\delta (t-t_{1})$ and that, for suitable $g(t)$, $\int_{0}^{t}g(t_{1})\delta
(t-t_{1})\,dt_{1}=\frac{1}{2}g(t)$, after some standard computations we
deduce that
\[
c_{j}(t)=e^{-\left( i\omega _{j}+\frac{2\pi \lambda _{inf}^{2}}{\Omega _{j}}%
\right) t}\times
\]%
\begin{equation}
\times \left[ \frac{1}{2}c_{j}(0)\left( 1+e^{\frac{2\pi \lambda _{inf}^{2}}{%
\Omega _{j}}t}\right) -\frac{1}{2}s_{j}(0)\left( e^{\frac{2\pi \lambda
_{inf}^{2}}{\Omega _{j}}t}-1\right) -i\lambda _{inf}\int_{\mathbb{R}%
}i_{j}(k)\eta _{j}^{(1)}(k,t)dk\right] .  \label{44}
\end{equation}%
Here we have defined the function
\[
\eta _{j}^{(1)}(k,t)=\frac{1}{i(\omega _{j}-\Omega _{j}\,k)+\frac{2\pi
\lambda _{inf}^{2}}{\Omega _{j}}}\left( \exp \left\{ \left( i(\omega
_{j}-\Omega _{j}\,k)+\frac{2\pi \lambda _{inf}^{2}}{\Omega _{j}}\right)
t\right\} -1\right) .
\]%
What we are interested in, as in the previous section, is, first of all, the
mean value of $\hat{K}_{j}(t)=c_{j}^{\dagger }(t)c_{j}(t)$ on a state $%
\left<.\right>$ over the whole system, i.e. a state over the traders and
their reservoirs. For each operator of the form $X_{sm}\otimes Y_{res}$, $%
X_{sm}$ being an operator of the stock market and $Y_{res}$ an operator of
the reservoir, we have
\[
\left\langle X_{sm}\otimes Y_{res}\right\rangle :=\left\langle \varphi _{%
\mathcal{G}},X_{sm}\varphi _{\mathcal{G}}\right\rangle \,\omega
_{res}(Y_{res}).
\]%
Here $\varphi _{\mathcal{G}}$ is defined in analogy with the vectors
introduced in the previous section, $\varphi _{\mathcal{G}}=\varphi
_{S_{1},K_{1},S_{2},K_{2}}$, while $\omega _{res}(.)$ is a state satisfying
the standard properties, \cite{bagbook},
\[
\omega _{res}(1\!\!1_{res})=1,\quad \omega _{res}(i_{j}(k))=\omega
_{res}(i_{j}^{\dagger }(k))=0,\quad \omega _{res}(i_{j}^{\dagger
}(k)i_{l}(q))=N_{j}^{(I)}(k)\,\delta _{j,l}\delta (k-q),
\]%
for a suitable function $N_{j}^{(I)}(k)$. Also, $\omega
_{res}(i_{j}(k)i_{l}(q))=0$, for all $j$ and $l$. Then we find
\begin{equation}
N_{K_{j}}(t):=\left\langle c_{j}^{\dagger }(t)c_{j}(t)\right\rangle =e^{-%
\frac{4\pi \lambda _{inf}^{2}}{\Omega _{j}}\,t}\times  \label{45}
\end{equation}%
\[
\times \left[ \frac{1}{4}N_{K_{j}}(0)\left( 1+e^{\frac{2\pi \lambda
_{inf}^{2}}{\Omega _{j}}t}\right) ^{2}+\frac{1}{4}N_{S_{j}}(0)\left( e^{%
\frac{2\pi \lambda _{inf}^{2}}{\Omega _{j}}t}-1\right) ^{2}+\lambda
_{inf}^{2}\int_{\mathbb{R}}N_{j}^{(I)}(k)|\eta _{j}^{(1)}(k,t)|^{2}dk\right]
,
\]%
where we have introduced, in analogy with $N_{K_{j}}$, $N_{S_{j}}(t):=\left%
\langle s_{j}^{\dagger }(t)s_{j}(t)\right\rangle $. Using now the fact that $%
s_{j}(t)=c_{j}(t)+e^{-i\omega _{j}\,t}(s_{j}(0)-c_{j}(0))$, we can also
deduce the time evolution of $N_{S_{j}}(t)$, which turns out to be
\[
N_{S_{j}}(t)=N_{K_{j}}(t)+e^{-\frac{2\pi \lambda _{inf}^{2}}{\Omega _{j}}%
\,t}\left( N_{S_{j}}(0)\left( 1-e^{\frac{2\pi \lambda _{inf}^{2}}{\Omega _{j}%
}\,t}\right) -N_{K_{j}}(0)\left( 1+e^{\frac{2\pi \lambda _{inf}^{2}}{\Omega
_{j}}\,t}\right) \right) +
\]%
\begin{equation}
+N_{K_{j}}(0)+N_{S_{j}}(0).  \label{46}
\end{equation}%
It is now easy to deduce the asymptotic behavior of the portfolio $\pi
_{j}(t)=N_{K_{j}}(t)+N_{S_{j}}(t)$. After some computation, and assuming
that $N_{j}^{(I)}(k)=N_{j}^{(I)}$ is constant in $k$, we deduce that
\[
\delta \pi _{j}:=\lim_{t\rightarrow \infty }\pi _{j}(t)-\pi _{j}(0)=-\frac{1%
}{2}\pi _{j}(0)+2\lambda _{inf}^{2}\Omega _{j}^{2}\,N_{j}^{(I)}\int_{\mathbb{%
R}}\frac{dk}{4\pi ^{2}\lambda _{inf}^{4}+\Omega _{j}^{2}(\omega _{j}-\Omega
_{j}k)^{2}}.
\]%
The integral can be computed using standard complex techniques, and we end
up with the following result
\begin{equation}
\delta \pi _{j}=-\frac{1}{2}\pi _{j}(0)+N_{j}^{(I)}.  \label{47}
\end{equation}%
In our idea, this conclusion is not particularly meaningful, since it states
that between $\tau _{1}$ and $\tau _{2}$, the one who is \emph{better
prepared}, is the one who starts with a smaller portfolio and for which the
associated reservoir has a larger value of $N_j^{(I)}$: if, for instance, $%
\pi _{1}(0)=\pi _{2}(0)$ and $N_{1}^{(I)}>N_{2}^{(I)}$, then $\delta \pi
_{1}>\delta \pi _{2}$.

What it is not very satisfying to us is the fact that, apparently, the
parameters of $H$ do not play any role in the behavior of the portfolios of
the traders, at least on a long time scale. This suggests that the model
should be improved further, and this is in fact the content of the next
section.

\section{The reservoir \protect\underline{generates} the information}

This section is devoted to a different, and probably more interesting model
where the reservoir, rather than being directly linked to the LoI, is used
\emph{to generate} the information reaching the traders. More in detail, the
Hamiltonian is

\begin{equation}
\left\{
\begin{array}{ll}
H=H_{0}+H_{int}, &  \\
H_{0}=\sum_{j=1}^{2}(\omega _{j}^{s}\hat{S}_{j}+\omega _{j}^{c}\hat{K}%
_{j}+\Omega _{j}\hat{I}_{j}+\int_{\mathbb{R}}\Omega _{j}^{(r)}(k)\hat{R}%
_{j}(k)\,dk), &  \\
H_{int}=\sum_{j=1}^{2}\left[ \lambda _{inf}\left( i_{j}(s_{j}^{\dagger
}+c_{j}^{\dagger })+i_{j}^{\dagger }(s_{j}+c_{j})\right) +\gamma _{j}\int_{%
\mathbb{R}}(i_{j}^{\dagger }r_{j}(k)+i_{j}r_{j}^{\dagger }(k))\,dk\right] ,%
\label{51} &
\end{array}%
\right.
\end{equation}%
where $\hat{R}_{j}(k)=r_{j}^{\dagger }(k)r_{j}(k)$, $\hat{S}_{j}$, $\hat{K}%
_{j}$ and $\hat{I}_{j}$ are defined as in Section III, and the following
CCRs are assumed,
\[
\lbrack s_{j},s_{k}^{\dagger }]=[c_{j},c_{k}^{\dagger
}]=[i_{j},i_{k}^{\dagger }]=1\!\!1\delta _{j,k},\quad \lbrack
r_{j}(k),r_{l}^\dagger(q)]=1\!\!1\delta _{j,l}\delta (k-q),
\]%
all the other commutators being zero. The reservoir is described here by the
operators $r_{j}(k)$, $r_{j}^{\dagger }(k)$ and $\hat{R}_{j}(k)$, and it is
used to model the set of all the rumors, news, and external facts which, all
together, create the final information. This Hamiltonian contains a free
\emph{canonical} part $H_{0}$, while the two contributions in $H_{int}$
respectively describe: (i) the same mechanism considered in Section III:
when the LoI increases, the value of the portfolio decreases and vice-versa;
(ii) the LoI increases when the "value" of the reservoir decreases, and
viceversa. Considering, for instance, the contribution $i_{j}r_{j}^{\dagger
}(k)$ in $H_{int}$ we see that the LoI decreases (so that the trader is
\emph{better informed}) when a larger amount of news, rumors, etc. reaches
the trader. Once again, no interaction between $\tau _{1}$ and $\tau _{2}$
is considered in (\ref{51}), since this is not the main aim of this paper.

As in the previous models, some self-adjoint operators are preserved during
the time evolution. These operators are $\hat M_j=\hat S_j+\hat K_j+\hat
I_j+\hat R_j=\hat \pi_j+\hat I_j+\hat R_j$, $j=1,2$. Here the only new
operator, with respect to those introduced in Section III, is $\hat
R_j=\int_{\mathbb{R}}r_j^\dagger(k)r_j(k)\,dk$. Then we can check that $%
[H,\hat M_j]=0$, $j=1,2$. This implies that what is constant in time is the
sum of the portfolio, the LoI and of the \emph{reservoir input} of each
single trader. Once again, there is no general need, and in fact it is not
required, for the cash or the number of shares to be constant in time.

The Heisenberg differential equations of motion can now be easily deduced:
\begin{equation}
\left\{
\begin{array}{ll}
\frac{d}{dt}s_{j}(t)=-i\omega _{j}^{s}s_{j}(t)-i\lambda _{inf}\,i_{j}(t), &
\\
\frac{d}{dt}c_{j}(t)=-i\omega _{j}^{c}c_{j}(t)-i\lambda _{inf}\,i_{j}(t), &
\\
\frac{d}{dt}i_{j}(t)=-i\Omega _{j}i_{j}(t)-i\lambda
_{inf}(s_{j}(t)+c_{j}(t))-i\gamma _{j}\int_{\mathbb{R}}r_{j}(k,t)\,dk &  \\
\frac{d}{dt}r_{j}(k,t)=-i\Omega _{j}^{(r)}(k)\,r_{j}(k,t)-i\gamma
_{j}\,i_{j}(t).\label{52} &
\end{array}%
\right.
\end{equation}%
In the previous section, to simplify the treatment, we required that $\omega
_{j}^{s}=\omega _{j}^{c}$. However, also in view of the results we have
deduced, we will avoid making this assumption now. We do not give the
details of the solution of this system here, details which can be deduced by
\cite{bagbook}. We just discuss the main steps. First of all, we rewrite the
last equation in its integral form:
\[
r_{j}(k,t)=r_{j}(k)e^{-i\Omega _{j}^{(r)}(k)t}-i\gamma
_{j}\int_{0}^{t}i_{j}(t_{1})e^{-i\Omega _{j}^{(r)}(k)(t-t_{1})}\,dt_{1},
\]%
and we replace this in the differential equation for $i_{j}(t)$. Assuming
that $\Omega _{j}^{(r)}(k)=\Omega _{j}^{(r)}\,k$, and proceeding as in the
previous section, we deduce that
\begin{equation}
\frac{d}{dt}i_{j}(t)=-\left( i\Omega _{j}+\frac{\pi \gamma _{j}^{2}}{\Omega
_{j}^{(r)}}\right) i_{j}(t)-i\gamma _{j}\int_{\mathbb{R}}r_{j}(k)e^{-i\Omega
_{j}^{(r)}\,kt}\,dk-i\lambda _{inf}(s_{j}(t)+c_{j}(t)).  \label{53}
\end{equation}%
In the rest of this section we will work under the assumption that the last
contribution in this equation can be neglected, when compared to the other
ones. In other words, we are taking $\lambda _{inf}$ to be very small.
However, our procedure is slightly better than simply considering $\lambda
_{inf}=0$ in $H$ above, since we will keep the effects of this term in the
first two equations in (\ref{52}). Solving now (\ref{53}) in its simplified
expression, and replacing the solution $i_{j}(t)$ in the first equation in (%
\ref{52}), we find:
\begin{equation}
s_{j}(t)=e^{-i\omega _{j}^{s}t}\left( s_{j}(0)-i\lambda _{inf}\alpha
_{j}(t)\,i_{j}(0)-\lambda _{inf}\gamma _{j}\int_{\mathbb{R}}r_{j}(k)\,\eta
_{2,j}(k,t)\,dk\right) ,  \label{54}
\end{equation}%
where we have defined
\[
\alpha _{j}(t)=\frac{e^{(i\omega _{j}^{s}-\Gamma _{j})t}-1}{i\omega
_{j}^{s}-\Gamma _{j}},\qquad \eta _{2,j}(k,t)=\int_{0}^{t}\eta
_{1,j}(k,t_{1})e^{(i\omega _{j}^{s}-\Gamma _{j})t_{1}}\,dt_{1},
\]%
with
\[
\Gamma _{j}=i\Omega _{j}+\frac{\pi \gamma _{j}^{2}}{\Omega _{j}^{(r)}}%
,\qquad \eta _{1,j}(k,t)=\frac{e^{(\Gamma _{j}-i\Omega _{j}^{(r)}k)t}-1}{%
\Gamma _{j}-i\Omega _{j}^{(r)}k}.
\]%
It is clear from (\ref{52}) that a completely analogous solution can be
deduced for $c_{j}(t)$. The only difference is that $\omega _{j}^{s}$ should
be replaced everywhere by $\omega _{j}^{c}$.

The states of the system extend those of the previous section: for each
operator of the form $X_{sm}\otimes Y_{res}$, where $X_{sm}$ is an operator
of the stock market and $Y_{res}$ an operator of the reservoir, we have
\[
\left\langle X_{sm}\otimes Y_{res}\right\rangle =\left\langle \varphi _{%
\mathcal{G}},X_{sm}\varphi _{\mathcal{G}}\right\rangle \,\omega
_{res}(Y_{res}).
\]%
Here $\varphi _{\mathcal{G}}$ is of the form $\varphi _{\mathcal{G}}=\varphi
_{S_{1},K_{1},I_{1},S_{2},K_{2},I_{2}}$, exactly as in Section III, while $%
\omega _{res}(.)$ is a state satisfying again
\[
\omega _{res}(1\!\!1_{res})=1,\quad \omega _{res}(r_{j}(k))=\omega
_{res}(r_{j}^{\dagger }(k))=0,\quad \omega _{res}(r_{j}^{\dagger
}(k)r_{l}(q))=N_{j}^{(r)}(k)\,\delta _{j,l}\delta (k-q),
\]%
for a suitable function $N_{j}^{(r)}(k)$, as in Section IV. Also, $\omega
_{res}(r_{j}(k)r_{l}(q))=0$, for all $j$ and $l$. Then $N_{S_{j}}(t)=\left%
\langle s_{j}^{\dagger }(t)s_{j}(t)\right\rangle $ assumes the following
expression:
\begin{equation}
N_{S_{j}}(t)=N_{S_{j}}(0)+\lambda _{inf}^{2}N_{I_{j}}(0)|\alpha
_{j}(t)|^{2}+\lambda _{inf}^{2}\gamma _{j}^{2}\int_{\mathbb{R}%
}N_{j}^{(r)}(k)|\eta _{2,j}(k,t)|^{2}\,dk,  \label{55}
\end{equation}%
where $N_{I_{j}}(0)=\left\langle i_{j}^{\dagger }(0)i_{j}(0)\right\rangle
=I_{j}$ and $N_{S_{j}}(0)=S_{j}$ are fixed by the quantum numbers of $%
\varphi _{\mathcal{G}}$. The expression for $N_{K_{j}}(t)=\left\langle
c_{j}^{\dagger }(t)c_{j}(t)\right\rangle $ is completely analogous to the
one above, with $\omega _{j}^{s}$ replaced by $\omega _{j}^{c}$, and the
portfolio of $\tau _{j}$, $\pi _{j}(t)$, is simply the sum of $N_{S_{j}}(t)$
and $N_{K_{j}}(t)$. What we are interested in, is the variation of $\pi
_{j}(t)$ over long time scales:
\[
\delta \pi _{j}:=\lim_{t,\infty }\pi _{j}(t)-\pi _{j}(0).
\]%
Formula (\ref{55}) shows that, if $\gamma _{j}$ is small enough, the
integral contribution is expected not to contribute much to $\delta \pi _{j}$%
. For this reason, we will not consider it in the rest of the section. We
now find
\begin{equation}
\delta \pi _{j}=\lambda _{inf}^{2}I_{j}(\Omega _{j}^{(r)})^{2}\left( \frac{1%
}{\pi ^{2}\gamma _{j}^{4}+(\omega _{j}^{s}-\Omega _{j})^{2}(\Omega
_{j}^{(r)})^{2}}+\frac{1}{\pi ^{2}\gamma _{j}^{4}+(\omega _{j}^{c}-\Omega
_{j})^{2}(\Omega _{j}^{(r)})^{2}}\right) .  \label{56}
\end{equation}%
Let us now recall that, at $t=0$, the two traders are equivalent: $\omega
_{1}^{c}=\omega _{2}^{c}=:\omega ^{c}$, $\omega _{1}^{s}=\omega
_{2}^{s}=:\omega ^{s}$, $\Omega _{1}^{(r)}=\Omega _{2}^{(r)}$ and the
initial conditions are $S_{1}=S_{2}$, $K_{1}=K_{2}$ and $I_{1}=I_{2}$. The
main difference between $\tau _{1}$ and $\tau _{2}$ is in $\Omega _{1}$
which is taken larger than $\Omega _{2}$: $\Omega _{1}>\Omega _{2}$\footnote{%
The case $\Omega _{1}<\Omega _{2}$ can be easily deduced, by exchanging the
role of $\Omega_1$ and $\Omega_2$.}. With this in mind, we will consider
three different cases: (a) $\gamma _{1}=\gamma _{2}$; (b) $\gamma
_{1}>\gamma _{2}$; (c) $\gamma _{1}<\gamma _{2}$. In other words, we are
allowing a different interaction strength between the reservoir and the
information term in $H$.

Let us consider {the }first situation (a): $\gamma _{1}=\gamma _{2}$ and $%
\Omega _{1}>\Omega _{2}$. In this case it is possible to check that $\delta
\pi _{1}<\delta \pi _{2}$, at least if $|\omega ^{c}-\Omega _{2}|<|\omega
^{c}-\Omega _{1}|$ and $|\omega ^{s}-\Omega _{2}|<|\omega ^{s}-\Omega _{1}|$%
. Notice that these inequalities are surely satisfied in our present
assumptions if $\Omega _{1}$ and $\Omega _{2}$ are sufficiently larger than $%
\omega ^{c}$ and $\omega ^{s}$. In this case the conclusion is, therefore,
that the larger the LoI, the smaller the increment in the value of the
portfolio. Needless to say, this is exactly what we expected to find in our
model. Exactly the same conclusion is deduced in case (b): $\gamma
_{1}>\gamma _{2}$ and $\Omega _{1}>\Omega _{2}$. In this case the two
inequalities produce the same consequences: we are \emph{doubling} the
sources of the LoI (one from $H_{0}$ and one from the interaction), and this
implies a smaller increment of $\pi _{1}$. Case (c): $\gamma _{1}<\gamma
_{2} $ and $\Omega _{1}>\Omega _{2}$, is different. In this case, while $%
H_{0}$ implies that $\tau _{1}$ is \emph{less informed} (or that the quality
of his information is not good enough), the inequality $\gamma _{1}<\gamma
_{2}$ would imply exactly the opposite. The conclusion is that, for fixed $%
\Omega _{1}$ and $\Omega _{2}$, there exists a critical value of $(\gamma
_{1},\gamma _{2})$ such that, instead of having $\delta \pi _{1}<\delta \pi
_{2}$, we will have exactly the opposite inequality, $\delta \pi _{1}>\delta
\pi _{2}$.

We should remind that these conclusions have been deduced under two
simplifying assumptions which consist in neglecting the last contributions
in (\ref{53}) and in (\ref{55}). Of course, to be more rigorous, we should
also have some control on these approximations. However, we will not do this
here.

\vspace{3mm}

As we see, this model is realistic and more than reasonable. Moreover, it
might be worth to stress that in all the models considered in this paper,
since the two traders do not interact with each other but only with the
information, there is absolutely no need to limit the system to a simple
two-traders stock market. In other words, as far as we are interested in the
preliminary phase of the market, the interval $[0,t_1]$ introduced in
Section II, we can easily extend all our models and our conclusions to
larger markets, with an arbitrarily large number of traders.

\section{Conclusions}

In this paper we have proposed several models to incorporate the role of the
information in a simplified, quantum-like, model of {a }stock market. In
particular, we have considered what happens before the traders begin to
interact, i.e. in a phase where the traders, identical at time $t=0$, begin
to experience some information coming from inside the system (Section III)
or from some \emph{surrounding world} (Sections IV and V). Each one of the
proposed models produce an interesting dynamical behavior, and the last one,
in particular, appears to be quite promising for a deeper analysis.

The natural Step 2 of our research would consist in the analysis of what
happens to the traders of a market prepared as, say, in Section V, when they
start to interact, i.e. to buy and sell shares. Of course, the natural
choice of the exchange Hamiltonian $H_{ex}$ introduced in Section II is the
following, see \cite{bagbook},
\[
H_{ex}=\nu \left( s_{1}^{\dagger }c_{1}s_{2}c_{2}^{\dagger }+s_{2}^{\dagger
}c_{2}s_{1}c_{1}^{\dagger }\right) ,
\]%
which describes the fact that $\tau _{1}$ buys a share from $\tau _{2}$, and
pays for that (the first term) or that the opposite happens (second term).
Notice that, in $H_{ex}$, we are implicitly assuming that the price of the
share is one. Of course, a more interesting model should also contain some
reasonable dynamics for the price of the shares. This is very hard, and it
is also part of our future plans.

{This paper also shows that by using tools from quantum mechanics we are
able to formalize information dynamics in a macroscopic setting. The work
presented here indicates that even with the use of such tools, the economic
intuition remains robust: i.e. the loss of information level affects the
incremental value of portfolios and this conclusion is maintained under the
scenarios that }$\gamma _{1}\geq \gamma _{2}$. {When interaction between
traders is set into action in a forthcoming paper, the role of i) the level;
and ii) the type of the interest rate will be able to be taken into account.
A related consequence of allowing for transactions between traders to occur,
will be to investigate how the loss of information can affect the potential
existence of arbitrage in transactions. Given that the (non) existence of
arbitrage plays such a fundamental role in the allowable use of the risk
free rate of interest and the pricing of assets, we may well be in a
position to specifically link the level of loss of information (maybe via a
treshold value) with the (non) existence of arbitrage. Hence, if such a
relationship were to exist, then extensions on the approach presented in
this paper, can provide for a proper vehicle to better model the concept of
arbitrage altogether.}

\section*{Acknowledgements}

F.B. acknowledges financial support from Universit\`a di Palermo.


\begin{thebibliography}{99}
\bibitem{bag1} F. Bagarello, \emph{An operatorial approach to stock markets}%
, J. Phys. A, \textbf{39}, 6823-6840 (2006)

\bibitem{bag2} F. Bagarello, \emph{Stock Markets and Quantum Dynamics: A
Second Quantized Description}, Physica A, \textbf{386}, 283-302 (2007)

\bibitem{bag3} F. Bagarello, \emph{Simplified Stock markets and their
quantum-like dynamics}, Rep. on Math. Phys., \textbf{63}, No. 3, 381-398
(2009)

\bibitem{bag4} F. Bagarello \emph{A quantum statistical approach to
simplified stock markets}, Physica A, \textbf{388}, 4397-4406 (2009)

%\bibitem{bag5} F. Bagarello, F. Oliveri, \emph{An operator-like description
%of love affairs}, SIAM J. Appl. Math., \textbf{70}, No. 8, 3235--3251 (2010)

%\bibitem{bag6} F. Bagarello, F. Oliveri \emph{Quantum Modeling of Love
%Affairs}, Proceedings of the XIV WASCOM 2009, Palermo, 7-14, World
%Scientific (2010)

%\bibitem{bag7} F. Bagarello, \emph{Damping in quantum love affairs}, Physica
%A, \textbf{390}, 2803-2811 (2011)

%\bibitem{bag8} F. Bagarello, \emph{Few simple rules to fix the dynamics of
%classical systems using operators}, Int. J. Theor. Phys., \textbf{51}, N. 7,
%2077-2085 (2012)

%\bibitem{bag9} F. Bagarello, F. Oliveri, \emph{An operator description of
%interactions between populations with applications to migration}, Math. Mod.
%and Meth. in Appl. Sci., in press

\bibitem{bagbook} F. Bagarello, \emph{Quantum dynamics for classical
systems: with applications of the Number operator}, Wiley Ed., New York,
(2012)

\bibitem{khren1} A. Yu. Khrennikov, \emph{Information dynamics in cognitive,
psychological, social and anomalous phenomena}, Kluwer, Dordrecht (2004)

\bibitem{hav1} E. Haven, \emph{The variation of financial arbitrage via the
use of an information wave function}, Int. J. Theor. Phys., \textbf{51},
193-199 (2008)

\bibitem{hav2} E. Haven, \emph{It\^{o}'s Lemma with quantum calculus: some
implications}, Found. Phys., \textbf{41, }No. 3, 529-537 (2010)

\bibitem{accbouk} L. Accardi, A. Boukas, \emph{The quantum Black-Scholes
equation}, Glob. J. Pure Appl. Math, \textbf{2}, No. 2, 155-170 (2006)

\bibitem{aerts0} D. Aerts, B. D'Hooghe, S. Sozzo, \emph{A quantum approach
to the stock market}, arXiv:1110.5350v1 [q-fin.GN]

\bibitem{atau} A. Ataullah, I. Anderson, M. Tippett, \emph{A wave function
for stock market returns}, Physica A, \textbf{388}, 455--461 (2009)

\bibitem{baa} B.E. Baaquie, \emph{Quantum Finance}, Cambridge University
Press, (2004)

\bibitem{baaphysrev} B. E. Baaquie, \emph{Interest rates in quantum finance:
the Wilson expansion}, Phys. Rev. E, \textbf{80}, 046119 (2009)

\bibitem{piotr} E. W. Piotrowski, J. S{\l }adkowski, \emph{Quantum diffusion
of prices and profits}, Physica A \textbf{345}, 185-195 (2005)

\bibitem{seg} W. Segal, I. E. Segal, \emph{The Black--Scholes pricing
formula in the quantum context}, Proc. Natl. Acad. Sci. USA, \textbf{95},
4072--4075 (1998)

\bibitem{manu} E. Haven, A. Khrennikov, \emph{Quantum social science},
Cambridge University Press, New York (2013)

\bibitem{chou} O. Choustova, \emph{Quantum Bohmian model for financial market%
}, Physica A, \textbf{374}, 304-314 (2007)

\bibitem{bohm1} D. Bohm, {\emph{A suggested interpretation of the
quantum theory in terms of hidden variables}. }Phys. Rev., \textbf{85},
166-179 (1952a)

\bibitem{bohm2} D. Bohm,{\ \emph{A suggested interpretation of the
quantum theory in terms of hidden variables}. }Phys. Rev.\textit{, }\textbf{%
85}\textit{, }180--193 (1952b)

\bibitem{rom} P. Roman, \emph{Advanced quantum mechanics}, Addison--Wesley,
New York (1965)

\bibitem{reg} M. Reginatto{, \emph{Derivation of the equations of
nonrelativistic quantum mechanics using the principle of minimum Fisher
information}. }Phys. Rev. A, 1775-1778 (1998)

\bibitem{hawkins} R. J. Hawkins, M. Aoki, B. J. Frieden{, \emph{%
Asymmetric information and macroeconomic dynamics}. }Physica A\textit{, }%
\textbf{389}, 3565-3571 (2010)
\end{thebibliography}
\end{document}